\documentclass[a4paper,11pt]{article}
\usepackage{jinstpub} 
\usepackage{lineno}
\usepackage{siunitx}
\usepackage{subcaption}

\title{\boldmath Development of CMOS LGAD sensors for the ALICE~3 Time of Flight detector}

\collaboration[c]{on behalf of ALICE3 Timing Layer group}

\author{U. Follo}
\affiliation{Dipartimento di Elettronica e Telecomunicazioni (DET) - Politecnico di Torino,\\ Corso Duca degli Abruzzi 24, 10129, Torino, Italy}
\affiliation{Istituto Nazionale di Fisica Nucleare (INFN) - Sezione di Torino,\\
Via P. Giuria 1, 10125, Torino, Italy}

\emailAdd{umberto.follo@polito.it}

\abstract{The next-generation ALICE 3 experiment at the High-Luminosity LHC (HL-LHC) requires detector technologies that combine fine spatial resolution, fast timing, and an extremely low material budget.. This paper presents the design, characterization, and beam-test performance of \textit{MadPix}, a monolithic CMOS sensor featuring an internal avalanche gain layer. The sensor is implemented in a 110 nm CMOS imaging process and demonstrates the portability of the Low Gain Avalanche Diode concept to a standard CMOS technology. The results showed an intrinsic gain between 10 and 13 and a time resolution of \SI{75}{ps}.}

\keywords{Front-end electronics for detector readout, Solid state detectors, Timing detectors}

\begin{document}
\maketitle
\flushbottom

\section{Introduction}

The ALICE experiment studies strongly interacting matter through high-energy nucleus–nucleus collisions. The forthcoming ALICE~3 upgrade \cite{Scoping}, which is planned to be installed for Run 5 (2036-2041), is a next-generation heavy-ion experiment that will be based on an innovative detector concept. It will fully exploit the increased luminosity of the Large Hadron Collider (HL-LHC), and it will perform precision measurements of rare probes. The experiment will rely on advanced silicon sensors with superb pointing resolution, excellent tracking, and Particle Identification (PID) over a large acceptance ($|\eta| < 4$). 

To meet the physics goals, the Time Of Flight (TOF) system used for PID needs \SI{20}{ps} time resolution across $\SI{45}{m^2}$ of total surface while minimizing material ($<3\% X_0$) and power ($<\SI{200}{mW/cm^2}$) budgets. In addition, the modest radiation environment ($<10^{13}\cdot 1\,$MeV $n_{eq}$/cm$^2$) and the low hit rate ($<\SI{280}{kHz/cm^2}$) allow for a monolithic CMOS sensor approach. However, the most mature technology for silicon timing sensors is the Low Gain Avalanche Diode (LGAD) \cite{Ferrero}, which offers excellent time resolution but requires hybridization, increasing cost and material. Monolithic approaches like MAPS, in contrast, offer low mass and low cost detectors but insufficient timing performance \cite{Rinella}.

The goal of the Monolithic CMOS Avalanche Detector PIXelated \textit{MadPix} prototype was to merge these two concepts by introducing an internal gain mechanism within a standard monolithic CMOS process, enabling state-of-the-art time resolution with scalable integration.

\section{CMOS-LGAD design and results}

In standard Monolithic Active Pixels Sensors (MAPS), charge is collected by drift or diffusion in a depleted epitaxial layer, with in-pixel electronics fabricated in the same substrate. While this architecture minimizes cost and simplifies production, the achievable timing resolution is limited to the nanosecond scale due to small signal amplitudes and slow diffusion components \cite{Rinella}. The starting point is the standard CMOS Image Sensor process developed by the ARCADIA collaboration, which allows for the complete depletion of the substrate and uniform electric fields using large collection electrodes \cite{Pancheri}.

\begin{figure}[!ht]
\centering
\includegraphics[width=0.5\linewidth]{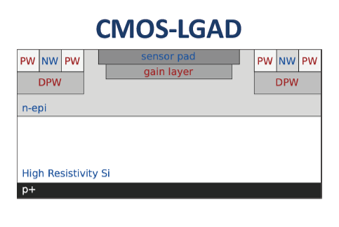}
\caption{Sensor cross-section of the CMOS-LGAD structure showing the n$^+$ sensor pad, the p$^+$ gain layer, the Deep P-Well (DPW) hosting PMOS and NMOS transistors, the active region (n-epi and High Resistivity Si), and the p$^+$ substrate.}
\label{fig:sensor_section}
\end{figure}

To improve the signal-to-noise ratio (SNR), the \textit{CMOS-LGAD} approach introduces a p$^+$-type gain layer just below the n$^{+}$ collection electrode as shown in Fig. \ref{fig:sensor_section}. This layer generates a localized high electric field (>\SI{300}{kV/cm}), causing controlled avalanche multiplication of drifting electrons. The resulting gain of 10–30 substantially enhances timing performance without an increase in the power consumption.

The device has been fabricated using LFoundry’s 110 nm CMOS imaging process, allowing full integration of the front-end electronics on the same substrate in the pixel borders. The active region is \SI{48}{\micro m} thick and can be fully depleted through backside biasing (from –20~V to –40~V). A separate topside voltage between 45 V and 65 V manages the gain.

\begin{figure}[!ht]
\centering
\includegraphics[width=0.8\linewidth]{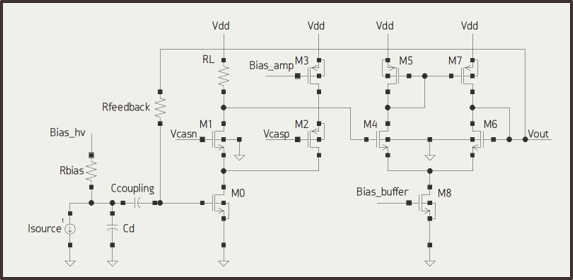}
\caption{Schematic of the MadPix front-end.}
\label{fig:schematic}
\end{figure}

The \textit{MadPix} demonstrator is designed with eight matrices of 64 pixels each, i.e., 512 pixels in total. The pixel size is $250\times100~\mu\text{m}^2$ and includes the front-end in two $250\times8~\mu\text{m}^2$ deep p-wells placed on top and bottom of the pixel.
The front-end is based on a cascoded common source followed by a differential amplifier used as a buffer (Fig. \ref{fig:schematic}). An additional buffer stage is placed outside the matrix and operates at 3.3 V. This stage is needed to drive the output pad and the PCB lines. Special attention was devoted to the design of the coupling capacitor between the sensor and the front end. This is a key element that AC couples the collector electrode, set at a voltage greater than 45 V, and the input transistor polarized between 0 and 1.2 V. The average front-end power consumption per channel is \SI{0.2}{mW}.
The prototype was fabricated in a dedicated ARCADIA Engineering Run (ER-2023) while the process tuning was carried out with two short-loop (SL-2024 and SL-2025). Engineering runs involve full mask set design and hence they come with a high degree of freedom, but at a high cost.
Instead, short loops reuse the same mask set fabricated in ER but modify implant doses or active thickness at reduced cost. The first ER in 2023 demonstrated the feasibility of gain layer integration. The following SL in 2024 targeted optimization of the gain implant dose.

Initial characterizations were carried out on structures without integrated electronics produced in ER-2023 using focused infrared laser illumination. The presence of intrinsic gain was measured, although lower than expected $\approx$3. To explain the very low gain, capacitance–voltage measurements between the collection electrode and the deep p-well were performed. A shallower profile was measured, and the TCAD parameters were tuned to match experimental results. A gain of 3 was extracted with the optimized simulation, hence a good agreement was established between simulation and data \cite{Follo}.
The substrate can be biased up to \SI{-28}{V} (punch-through point), while the breakdown voltage was measured above \SI{68}{V}, confirming the operability of the prototype.

A first beam test was performed at the CERN Proton Synchrotron (PS) using a proton beam at 10 GeV/c, in which the gain and time resolution were measured.
The ER-2023 prototypes exhibited a time resolution of \SI{120}{ps} for a power consumption of 0.17 mW/ch. For a gain of 3, the time resolution is dominated by electronic jitter as can be extracted from \cite{Follo}. Increasing the front-end power to 0.28 mW/ch, the jitter term is reduced, and \SI{105}{ps} can be achieved.

\begin{figure}[!ht]
\centering
\includegraphics[width=0.45\linewidth]{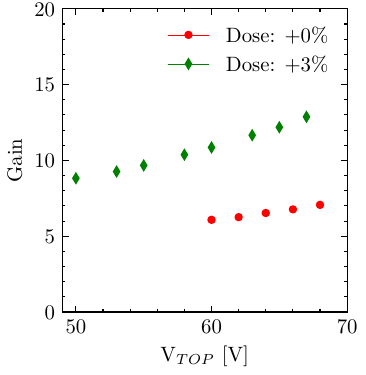}
\caption{Measured gain for SL-2024 for two different dose splits. In red, the reference dose (+ 0\%), while in green, a dose increase of + 3\%.}
\label{fig:gain}
\end{figure}

The SL-2024 was submitted using TCAD profile calibrated with ER-2023 and tuned to achieve a gain in the 10–20 range. Two dose splits were measured, and the gain results are shown in plot \ref{fig:gain}. The gain is evaluated as the ratio between the Most Probable Value (MPV) of the Landau distribution from a MadPix with internal gain and that from one without the gain-layer implant. The measured gain between 11 and 13 obtained with the +3\% dose is consistent with the design goal.
The combined sensor plus electronics time resolution is of \SI{88}{ps}, corresponding to an intrinsic sensor contribution of approximately \SI{75}{ps} obtained by subtracting the jitter term. This result was obtained at $V_\text{BACK}=-25$ V and $V_\text{TOP}=67$ V using a constant fraction algorithm (50\% amplitude) explained in \cite{Carnesecchi}.

\begin{figure}
\centering
\begin{subfigure}{0.490\textwidth}
        \centering
	\footnotesize
        \includegraphics[width=\linewidth]{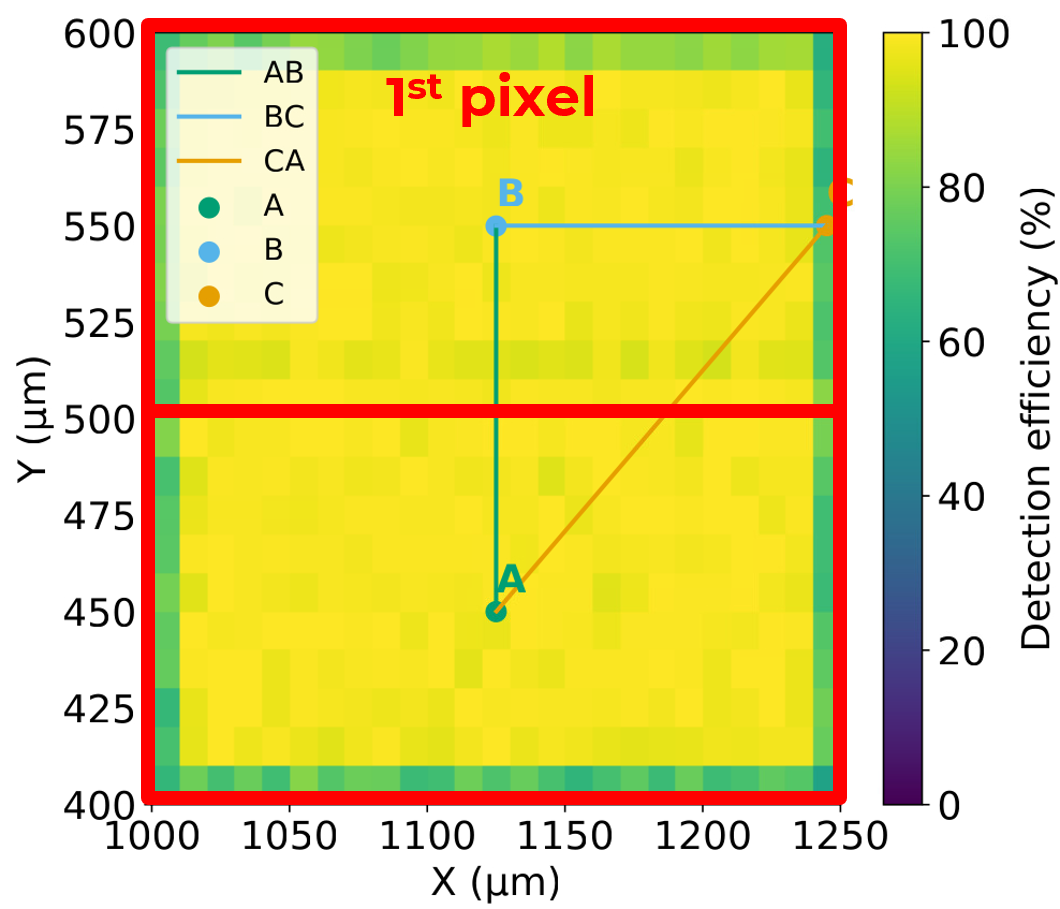}
        \caption{}
        \label{fig:eff_65}
\end{subfigure}
\hfill
\begin{subfigure}{0.490\textwidth}
        \centering
	\footnotesize
        \includegraphics[width=\linewidth]{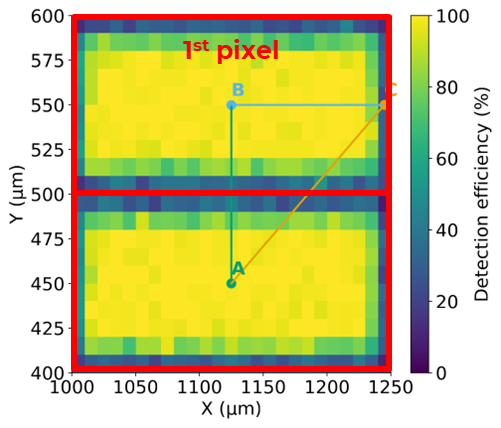}
        \caption{}
        \label{fig:eff_45}
\end{subfigure}
\caption{
Efficiency map extracted for 65 V (a) and 45 V (b) applied on the collection electrode, while biasing the substrate at -25 V. No efficiency drop between pixels is observed for (a), but it is present for (b).}
\end{figure}

\begin{figure}
    \centering
    \begin{subfigure}{0.490\textwidth}
        \centering
	\footnotesize
        \includegraphics[width=\linewidth]{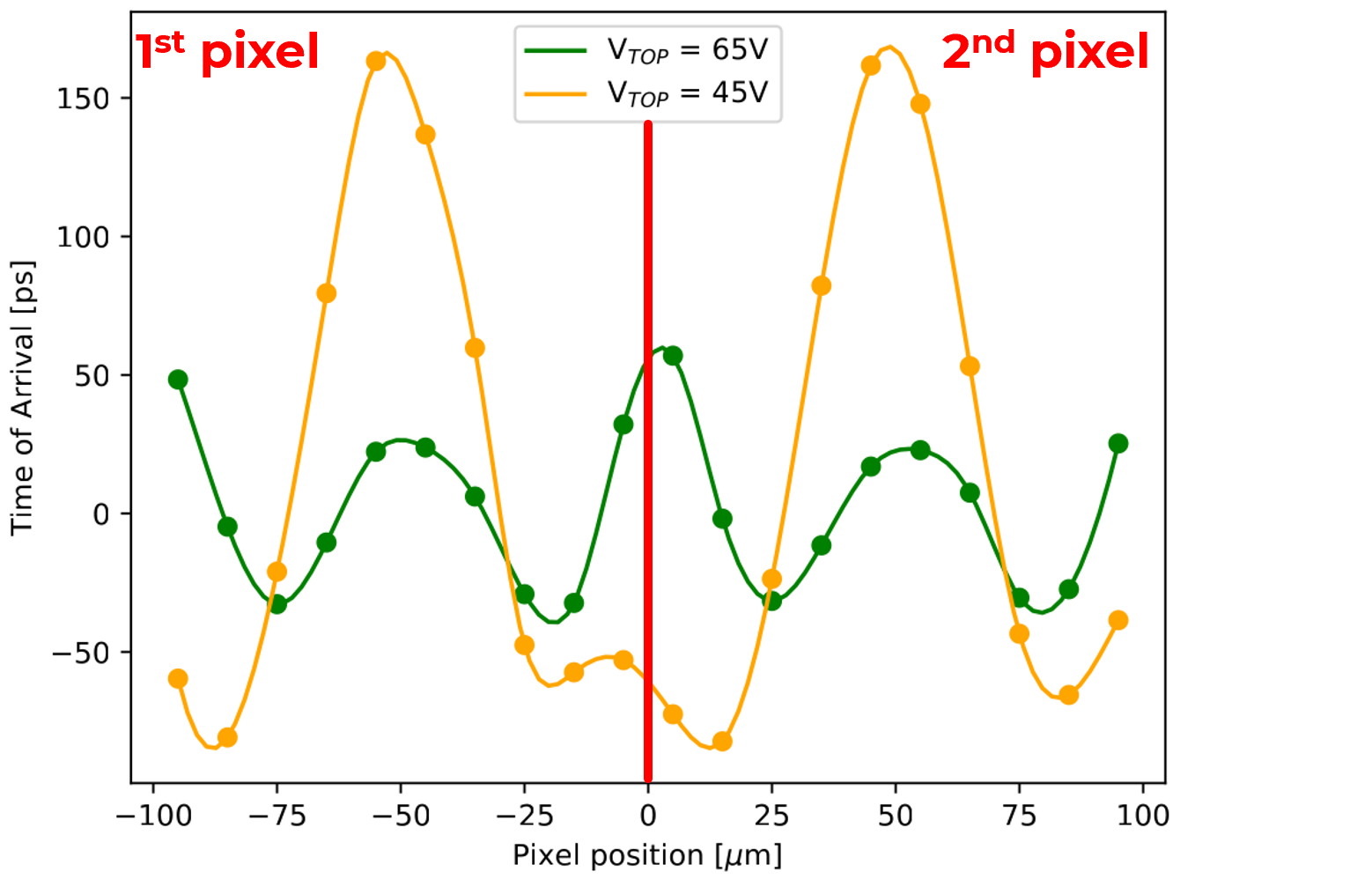}
        \caption{}
        \label{fig:toa}
    \end{subfigure}
    \hfill
    \begin{subfigure}{0.49\textwidth}
        \centering
	\footnotesize
        \includegraphics[width=0.91\linewidth]{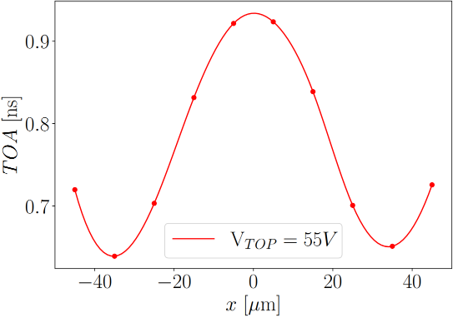}
        \caption{}
        \label{fig:toa_sim}
    \end{subfigure}
\caption{In-pixel Time Of Arrival map extracted at 65 V and 45 V from the DESY test beam data (a). The scan is performed on the 2 pixels acquired during the tests. TOA extracted from Monte Carlo simulations using TCAD profiles matched with data for one pixel at 55 V (b). The substrate is biased at -25 V in all the configurations.}
\end{figure}

An additional beam test was planned at the DESY Test Beam Facility (5 GeV/c electrons), where the DESY group allows for the use of a six MIMOSA planes telescope providing \SI{3.5}{\micro m} tracking resolution \cite{Desy} evaluated with the telescope optimizer inserting the material budget \cite{Telescope}. The Device Under Test (DUT) was synchronized with the telescope through the AIDA TLU and EUDAQ2 data acquisition system.
The telescope is required in order to study efficiency, time resolution, gain, and arrival time as a function of the particle impinging position. Two adjacent pixels were analyzed and at an operating threshold of 50 mV and $V_\text{TOP}=65$ V, the total pixel detection efficiency reached 95 \%. The efficiency map is shown in Fig. \ref{fig:eff_65}. The edge inefficiencies can be attributed to reconstruction uncertainties rather than charge loss. Indeed, between pixels (Y=500), no efficiency drop is present. Instead, at lower voltages (Fig. \ref{fig:eff_45}), a drop is present and the resolution goes down to 20\%.
Due to the low electrons rate (100 Hz), the statistic does not allow for a time resolution or time of arrival map, as for the efficiency. However, a cut in the X dimension can be applied between 1050 and 1150 $\mu m$ in Fig. \ref{fig:eff_65} and only the Y axis is binned. The Time-Of-Arrival (TOA) behavior is displayed in Fig. \ref{fig:toa} and reveals huge non-uniformities along the short pixel dimension at low gain. At $V_\text{TOP}=45$ V, the electric field minimum in the pixel center induces a delayed signal, while at $V_\text{TOP}=65$ V the field became more uniform, yielding a TOA spread below \SI{50}{ps} across the pixel. The observations were validated with TCAD-driven Monte-Carlo simulations using 180 GeV pion tracks as shown in Fig. \ref{fig:toa_sim}. The TOA can be used as an electric field probe inside the substrate because it can show where and if the carrier velocity is saturated.

Performing the same study carried out for the TOA, the time resolution reaches sub-\SI{70}{ps} in the pixel center. If the electronics jitter is subtracted in quadrature, the sensor time resolution drops to around 60 ps, which is the Landau contribution extracted from MC simulations.

\section{Conclusion and Outlook}

The results demonstrate the successful integration of the gain in a standard monolithic CMOS process with in-pixel front-end electronics. By tuning the simulation with a first engineering run, with a measured gain above 10. The timing performance (\SI{75}{ps}) represents a significant improvement over previous monolithic designs and establishes a path toward sub-\SI{20}{ps} resolution.

A second short-loop production was submitted in mid-2025 and will produce sensors with an active thickness of 15 $\mu$m. With these substrates, a timing resolution approaching \SI{30}{ps} is expected from Monte Carlo simulations.

The current pixel pitch is not optimized for timing due to design-rule constraints for this engineering run, and hence can not achieve the 20 ps goal. However, larger pixels are foreseen in later iterations to mitigate distortion effects and field non-uniformities. An engineering run planned for 2027 will validate the CMOS-LGAD concept on a full-reticle size chip with an optimized pixel pitch.

\acknowledgments

The author acknowledges the invaluable contributions of the members of the PS Group and the DESY Group for their support during test-beam activities. The measurements were performed at the Test Beam Facility at DESY Hamburg, a member of the Helmholtz Association (HGF).

\paragraph{Funding} This work received funding from the European Union’s Horizon Europe research and innovation programme under grant agreement No.,101057511.




\end{document}